\documentclass{osa-article}

\journal{oe}

\articletype{Research Article}

\usepackage{lineno}
\usepackage{bm, diffcoeff}
\usepackage{siunitx}[=v2]
\usepackage{algorithm, algpseudocode}

\newcommand{\Fxyop}{\mathcal{F}_{xy}}
\newcommand{\iFxyop}{\mathcal{F}_{xy}^{-1}}
\newcommand{\Fxybracket}[1]{\hspace{-0.1em}\left\{\hspace{-0.1em}#1\hspace{-0.1em}\right\}}
\newcommand{\Fxysbracket}[1]{\{#1\}}
\newcommand{\Fxy}[1]{\Fxyop\Fxybracket{#1}}
\newcommand{\iFxy}[1]{\iFxyop\Fxybracket{#1}}
\newcommand{\Fxys}[1]{\Fxyop\Fxysbracket{#1}}
\newcommand{\iFxys}[1]{\iFxyop\Fxysbracket{#1}}
\newcommand{\Loss}{\mathcal{L}}
\newcommand{\Grad}{\mathcal{G}}
\newcommand{\vPhi}{\bm{\Phi}}
\newcommand{\vH}{\mathbf{H}}
\newcommand{\vP}{\mathbf{P}}
\newcommand{\vQ}{\mathbf{Q}}
\newcommand{\vF}{\operatorname{F}}
\renewcommand{\Re}{\operatorname{Re}}
\algnewcommand\INPUT{\item[\textbf{Input:}]}
\algnewcommand\OUTPUT{\item[\textbf{Output:}]}
\newcommand{\vPhifzc}{\varphi_\mathrm{out}}
\newcommand{\vIssnp}{I_\mathrm{out}}
\newcommand{\cvarx}{v}
\newcommand{\cvary}{w}
\newcommand{\cvarz}{u}

\begin{document}

\title{High-fidelity intensity diffraction tomography with a non-paraxial multiple-scattering model}

\author{Jiabei Zhu,\authormark{1} Hao Wang,\authormark{1} and Lei Tian\authormark{1,2,*}}

\address{
  \authormark{1}Department of Electrical and Computer Engineering, Boston University, MA 02215, USA\\
  \authormark{2}Department of Biomedical Engineering, Boston University, MA 02215, USA\\
}

\email{\authormark{*}leitian@bu.edu} 



\begin{abstract}
  We propose a novel intensity diffraction tomography (IDT) reconstruction algorithm based on the split-step non-paraxial (SSNP) model for recovering the 3D refractive index (RI) distribution of multiple-scattering biological samples.
  High-quality IDT reconstruction requires high-angle illumination to encode both low- and high- spatial frequency information of the 3D biological sample.
  We show that our SSNP model can more accurately compute multiple scattering from high-angle illumination compared to paraxial approximation-based multiple-scattering models.
  We apply this SSNP model to both sequential and multiplexed IDT techniques. 
  We develop a unified reconstruction algorithm for both IDT modalities that is highly computationally efficient and is implemented by a modular automatic differentiation framework.
  We demonstrate the capability of our reconstruction algorithm on both weakly scattering buccal epithelial cells and strongly scattering live \textit{C. elegans} worms and live \textit{C. elegans} embryos.
\end{abstract}

\section{Introduction}
%

3D quantitative phase imaging (QPI) has found use in many biological applications by providing 3D refractive index (RI) information of the samples~\cite{Park2018}.
Optical diffraction tomography (ODT)~\cite{Jin17} is the most widely used technique for 3D QPI\@.
It works by directly measuring the complex field using an interferometry setup from multiple angles and then recovering the 3D RI distribution with an inverse scattering model.
However, the need for coherent illumination and an interferometric path complicates the system and makes it prone to speckle artifacts~\cite{Kim2014}.

\begin{figure}[ht!]
  \centering
  \includegraphics[width=0.90\textwidth]{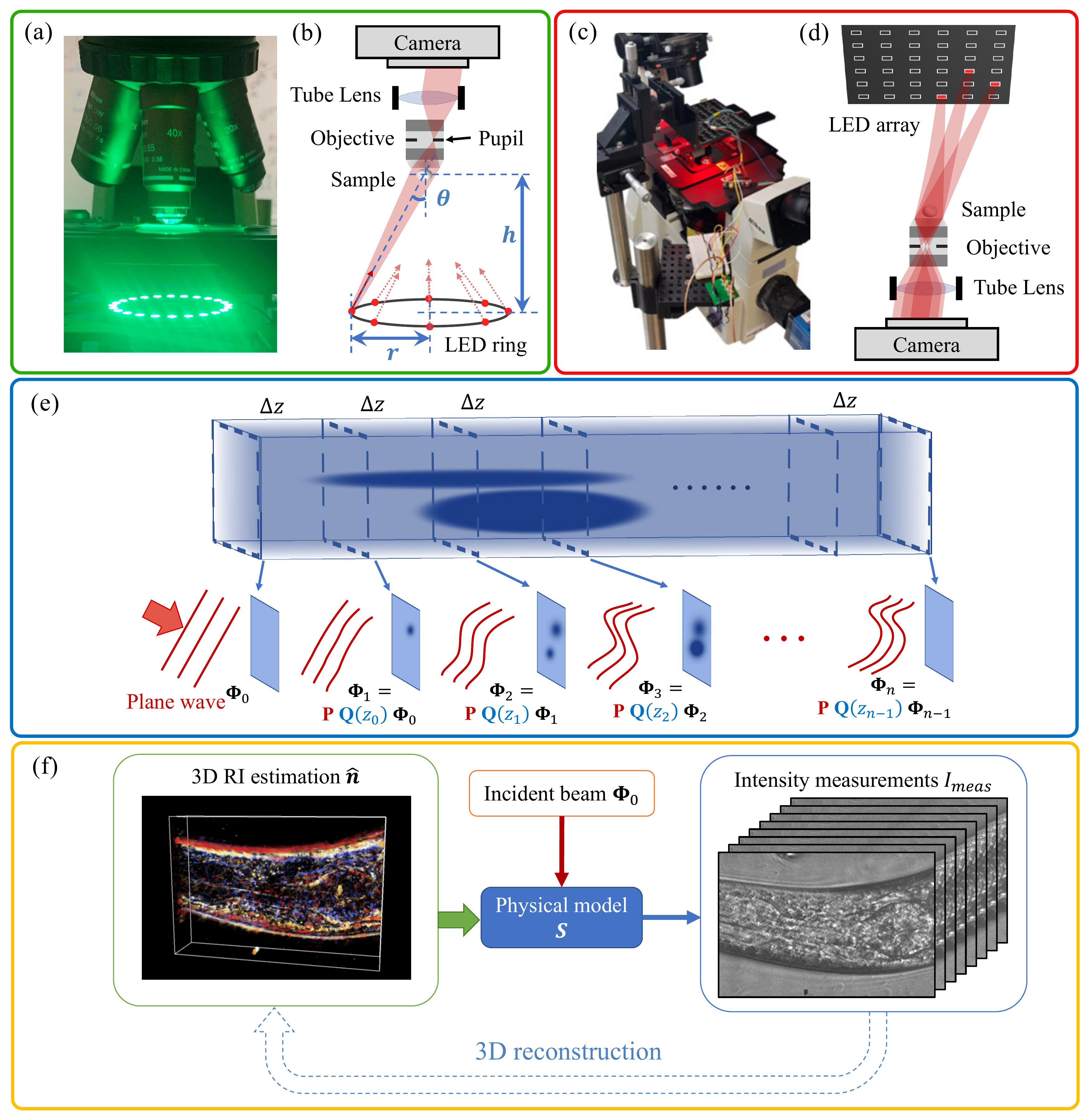}
  \caption{
    (a-b) Our annular IDT setup uses an LED ring of radius $r$ to illuminate the sample sequentially.
      The ring is placed $h$ away from the sample and the illumination angle $\theta=\arctan(r/h)$ matches the objective NA\@.
    (c-d) Our multiplexed IDT setup uses an LED matrix and illuminates the sample using several LEDs simultaneously.
    (e) The SSNP model involves successive propagation $\vP$ and scattering $\vQ$ operations to compute the scattered field and its axial derivative $\vPhi_i$ from one slice to the next.
    (f) 3D reconstruction by iteratively solving the SSNP-based optimization problem.
  }\label{fig1}
\end{figure}

Intensity diffraction tomography (IDT) is an alternative 3D QPI technique that recovers 3D phase information from intensity-only measurements on a ``non-interferometry'' setup~\cite{Gbur:02}.
Recently, several groups have demonstrated that IDT can be easily implemented on a standard microscope using a programmable LED array~\cite{Ling:18,Tian2015,Li2019,Chen:20,Zhou:20,Matlock2019,Li2022,odtinline2022,Horstmeyer:16}.
Most of the IDT acquisition strategies require taking hundreds of images to achieve high-resolution 3D RI reconstruction.
Recently, our group has developed two strategies to push the acquisition speed and enabled visualizing dynamic biological samples.
The annular IDT achieved a $\SI{10}{\hertz}$ volume rate by using an LED ring illuminator that provides illumination angles matching the objective's numerical aperture (NA)~\cite{Li2019}. 
The multiplexed IDT provided a $4 - \SI{10}{\hertz}$ acquisition rate by simultaneously tuning on multiple LEDs using the most widely used LED matrix~\cite{Matlock2019}.

The 3D reconstruction of IDT requires an inverse scattering model, similar to ODT\@.
The single-scattering models are based on the first-Born~\cite{Ling:18,Li2019,Matlock2019} or first-Rytov~\cite{odtinline2022} approximation, allowing closed-form solutions due to the linear relationship between the scattering potential and the measurements.
However, the accuracy of these models are fundamentally limited by the weak scattering assumption.
Recently, several multiple-scattering models have been developed based on the beam propagation method (BPM)~\cite{Tian2015,Chowdhury2019} and multi-layer Born (MLB) model~\cite{Chen:20} and showed enhanced resolution and fidelity on strongly scattering biological samples, such as \textit{C. elegans} embryos and worm.
However, the accuracy of the BPM degrades for high-NA systems since it relies on the paraxial approximation~\cite{Kamilov2016}, which limits its effectiveness for high-resolution imaging using high-NA optics.
The MLB model relies on the local weak scattering approximation within each layer, which limits the model fidelity for strongly scattering samples.
Recently, a split-step non-paraxial (SSNP) multiple-scattering model has been developed for ODT~\cite{Lim2019}, which demonstrated substantial improvement in reconstruction quality compared to the single-scattering and BPM methods.
This ODT setup, however, needs a laser and a spatial light modulator to provide the coherent illumination and a interferometric path to retrieve the phase information of the light field.
Besides, the large number of measurements used limits this system's ability of high-speed imaging for dynamic samples.
Inspired by this work, here we develop an SSNP IDT model for the two fast-acquisition IDT modalities.
We demonstrate that our model can effectively provide high-quality 3D RI reconstruction on several biological samples with simple ``non-interferometric'' IDT setups.

The SSNP model is a wave propagation model derived from the Helmholtz equation~\cite{Sharma04}.
Similar to the BPM and MLB, the SSNP works by axially splitting the sample into multiple slices and propagate the wave slice-wise.
The BPM and MLB rely on local homogeneous RI and weak RI contrast, respectively, to decouple the diffraction from the local phase modulation / single scattering within each slice.
Instead, the SSNP simultaneously propagates both the field and the axial derivative of the field so that the coupling between the diffraction and local phase modulation is modeled jointly without making assumptions about the local RI distribution~\cite{Lim2019}.
Due to the same ``multi-slice'' structure as the BPM and MLB, together with the fast Fourier transform (FFT)-based propagation implementation~\cite{Lim2019}, the SSNP incurs similar computational costs while providing improved fidelity on high-scattering biological samples.

Here we first extend the SSNP model to IDT and apply it to two IDT modalities, including the sequential and multiplexed IDT (as summarized in Fig.~\ref{fig1}a-d).
Next, we derive an efficient SSNP-IDT forward model (as illustrated in Fig.~\ref{fig1}e).
We then derive a reconstruction algorithm by solving a regularized least-squares optimization (Fig.~\ref{fig1}f).
We elucidate on how the analytical gradient of the SSNP-IDT model can be efficiently computed, akin to ``back-propagation''.
Furthermore, we devise a unified algorithmic framework that allows fast 3D recovery from both sequential and multiplexed IDT measurements based on a memory efficient automatic differentiation framework.
To quantitatively evaluate the fidelity of the forward model and reconstruction algorithm, we conduct numerical simulations.
Lastly, we experimentally demonstrate high-quality, large field-of-view (FOV) 3D RI reconstructions on buccal epithelial cells and a live \textit{C. elegans} worm from annular IDT and live \textit{C. elegans} embryos from  multiplexed IDT\@.

\section{Theory}
\subsection{SSNP theory}
The SSNP model first discretizes the 3D sample into a series of axial ($z$) slices and then calculates the internal field slice-by-slice ($xy$), as illustrated in Fig.~\ref{fig1}(e).
In the following, we provide the key steps to derive the governing equations for light propagating through the sample by following~\cite{Sharma04,Lim2019}, and then describe the integration to specific IDT systems in separate sections.

The scalar wave propagation in an inhomogeneous medium with a 3D RI distribution $n(\bm{r})$ satisfies the Helmholtz equation
\begin{equation}\label{hmhz}
  \left(\frac{\partial^2}{\partial x^2}+\frac{\partial^2}{\partial y^2}+\frac{\partial^2}{\partial z^2}\right)\,\varphi(\bm{r}) +
\,k_0^2 \; n^2(\bm{r}) \, \varphi(\bm{r}) = 0,
\end{equation}
where $\bm{r}=(x,y,z)$ denotes the 3D spatial position, $\varphi$ is the field and $k_0=2\pi/\lambda$ is the wavenumber in free space.
We derive the SSNP model by first rewriting Eq.~\eqref{hmhz} into a matrix form
\begin{equation}\label{diff}
  \diffp{\vPhi(\bm{r})}z=\bm{\mathrm{H}}(\bm{r})\;\vPhi(\bm{r}),
\end{equation}
where
\begin{equation}
  \label{eq:SSNP2}
\vPhi(\bm{r})=\begin{pmatrix}\varphi(\bm{r})\\ \frac{\partial \varphi}{\partial z}\end{pmatrix},
\ \vH(\bm{r})=\begin{pmatrix}
  0&1\\-\diffp[2]{}x-\diffp[2]{}y-\,k_0^2  n^2(\bm{r})\!\!&0
  \end{pmatrix}.
\end{equation}
Here the vector $\vPhi(\bm{r})$ contains both the field itself and its $z$-derivative, which is the quantity of interest in the SSNP-based wave propagation.
For brevity, we will refer $\vPhi$ to the 2D ``field'' at a slice in the rest of this article.
Next, we decompose the operator
$\bm{\mathrm{H}}(\bm{r}) = \vH_1 + \vH_2(\bm{r})$, which decouples the diffraction operator $\bm{\mathrm{H}}_1$
and the scattering operator $\bm{\mathrm{H}}_2(\bm{r})$:
\begin{equation}
  \label{eq:SSNP3}
  \vH_1=
  \begin{pmatrix}
    0&1\\-\diffp[2]{}x-\diffp[2]{}y\!-\!\,k_0^2 n_0^2\!&0
  \end{pmatrix},\,
  \vH_2(\bm{r})=
  \begin{pmatrix}
    0&0\\k_0^2(n_0^2-n^2(\bm{r}))\!&0
  \end{pmatrix}.
\end{equation}
Here $n_0$ is the RI of the background medium.
$\bm{\mathrm{H}}_1$ is spatially invariant describing the diffraction in the homogeneous background medium.
$\bm{\mathrm{H}}_2(\bm{r})$ is spatially variant based on the distribution of the scattering potential $\propto(n_0^2-n^2(\bm{r}))$, akin to the Born scattering model.
Equations~\eqref{diff}--\eqref{eq:SSNP3} constitute the governing equations in the SSNP model.

Next, we describe the numerical method to efficiently calculate Eq.~\eqref{diff}--\eqref{eq:SSNP3}.
First, we discretize the sample volume as a series of $z$-slices.
Second, to compute the field propagating a small distance $\Delta z$ between two adjacent slices, we approximate Eq.~\eqref{diff} as a first-order homogeneous linear system of differential equations with constant coefficients.
The solution is approximated as
\begin{equation}\label{ms}
  \vPhi_{xy}(z+\Delta z)\approx\exp(\vH(\bm{r})\Delta z)\,\vPhi_{xy}(z)\approx\vP\,\vQ(z)\,\vPhi_{xy}(z),
\end{equation}
where $\vPhi_{xy}(z)$ represents the 2D field at the axial position $z$, and
\begin{equation}\label{pq}
  \vP = \exp(\vH_1\Delta z),
  \ \vQ(z) = \exp(\vH_2(x,y,z)\Delta z).
\end{equation}

The operator $\vP$ computes the propagation by $\Delta z$ in the homogeneous background medium and can be efficiently computed by FFT as
\begin{align} \label{eq:FFT1}
 \vP\,\vPhi_{xy}
  &=\iFxys{\widetilde{\vP}\,\Fxys{\vPhi_{xy}}} \nonumber\\
  &=\iFxy{
    \begin{pmatrix}
      \cos(k_z \Delta z)&\sin(k_z \Delta z)/k_z\\
      -k_z\sin(k_z \Delta z)\!&\cos(k_z \Delta z)
    \end{pmatrix}
    \Fxys{\vPhi_{xy}}
  },
\end{align}
where $\Fxys{\cdot}$ and $\iFxys{\cdot}$ denote the 2D Fourier and inverse Fourier transform on the XY plane, respectively, and $k_x$ and $k_y$ are the corresponding $\mathbf{k}$ (wave vector) components. $k_z=\sqrt{k_0^2\,n_0^2-k_x^2-k_y^2}$ if $k_x^2+k_y^2 \leq k_0^2\,n_0^2$; otherwise, $\widetilde{\vP}=\bm{0}$ that removes the evanescent component.

The operator $\vQ(z)$ computes the scattering induced by the RI difference in the object slice at $z$ and is computed directly in the real space as
\begin{equation} \label{eq:Q}
  \vQ(z)\,\vPhi_{xy}=\begin{pmatrix}
    1&0\\
    k_0^2\left(n_0^2-n_{xy}^2(z)\right)\Delta z&1
  \end{pmatrix}\vPhi_{xy}.
\end{equation}
The SSNP model computes the exit field at slice $z_n$ by recursively applying $\vP$ and $\vQ$ slice-wise starting from the illumination field, as illustrated in Fig.~\ref{fig1}(e).
Detailed derivations for Eq.~\eqref{ms}--\eqref{eq:Q} are provided in Supplement 1.

\subsection{SSNP-based sequential IDT forward model}
The sequential IDT, such as annular IDT~\cite{Li2019}, uses a single LED to illuminate the sample in each measurement.
To derive the SSNP-based forward model, we first compute the illumination field and its axial derivative at $z=z_0$ as the initial condition to the SSNP equations.
Each illumination field is a plane wave $\varphi(\bm{r})=\varphi_0 \exp[j(k_x^{in}\,x + k_y^{in}\,y + k_z^{in}\,z)]$ where $\bm{k}^{in} = (k_x^{in}, k_y^{in}, k_z^{in})$ is the wave vector.
By omitting the constant factors $\varphi_0$ and $\exp(j\,k_z^{in}\,z_0)$, we have
\begin{equation}
  \vPhi_0=\exp[j(k_x^{in}\,x + k_y^{in}\,y)]\ 
    \begin{pmatrix}1\\j\,k_z^{in}\end{pmatrix}.
\end{equation}
For non-plane wave illumination, Supplement 1 provides a general treatment to construct $\vPhi_0$.

By propagating the illumination field through the sample with the SSNP equations, we obtain the exit field from the sample at plane $z_n$.
To compute the field through the microscope reaching the camera $\vPhi_\mathrm{cam}$, we first back-propagate the exit field from $z_n$ to the front focal plane ($z_\mathrm{focal}$), and then filter it by the microscope's pupil function:
\begin{equation} \label{eq:PhiCam}
  \vPhi_\mathrm{cam}=\mathcal{P}_\mathrm{NA}\vP_{\Delta z_f}\vPhi_{xy}(z_n),
\end{equation}
where $\vP_{\Delta z_f}=\exp(\vH_1\Delta z_f)$ is the propagation operator by a distance $\Delta z_f=z_\mathrm{focal}-z_n$.
$\mathcal{P}_\mathrm{NA}$ denotes the low-pass filtering by the pupil function, and can be computed efficiently in the Fourier space as a multiplication between the pupil function $\widetilde{\mathcal{P}_\mathrm{NA}}$ and the Fourier transform of the back-propagated field (and its axial derivative).
In our implementation, we assume a binary pupil function with a cutoff frequency at $k_0n_0\mathrm{NA}$ and ignore the apodization and aberrations.
The field $\vPhi_\mathrm{cam}$ contains both forward-propagating and back-propagating components. 
In practice, the back-propagating component will introduce high-frequency artifacts~\cite{Lim2019}, we extract only the forward-propagating component from $\vPhi_\mathrm{cam}$ by an operator $\vF$.
The forward-propagating field component at the camera plane $\vPhifzc$ is 
%
%
\begin{align}
  \nonumber \vPhifzc&=\vF\vPhi_\mathrm{cam}
  =\iFxy{
    \left(\Fxy{\varphi_\mathrm{cam}}-\frac{j}{k_z}\Fxy{\diffp{\varphi_\mathrm{cam}}z}\right)\middle/2\,
  }\\
  &=\iFxy{
    \left(\frac{1}{2},\ -\frac{j}{2 k_z}\right)
  \Fxys{\vPhi_\mathrm{cam}}}.
\end{align}
Detailed derivation of $\vF$ and a discussion about the bi-directional property of the SSNP model are provided in Supplement 1.
Finally, the intensity from sequential IDT estimated by the SSNP model is $\vIssnp = \left|\vPhifzc\right|^2$.

To summarize, our SSNP-based sequential IDT (SSNP-sIDT) forward model can be written in the following compact form
\begin{equation}\label{fwdfinal}
  \vIssnp = \left| \vF \mathcal{P}_\mathrm{NA} \vP_{\Delta z_f} \vP\vQ(z_{n-1}) \ldots \vP\vQ(z_1) \vP\vQ(z_0) \vPhi_{xy}(z_0) \right|^2.
\end{equation}

\subsection{SSNP-based multiplexed IDT forward model}
Multiplexed IDT uses several LEDs to illuminate the sample in each measurement~\cite{Matlock2019}.
Since the light from different LEDs are incoherent with each other, we can model the scattering process independently and compute the measured intensity as the sum of the intensities from all the LEDs in each measurement.
Therefore, the forward model of the SSNP-based multiplexed IDT (SSNP-mIDT) is similar to the sequential IDT but with an additional incoherent sum over Eq.~\eqref{fwdfinal}: 
\begin{equation}
  \vIssnp=\sum_{m=1}^M \vIssnp^m,
\end{equation}
where $\vIssnp^m$ is the intensity from the $m$th LED\@.
Each multiplexed measurement uses $M$ LEDs.

\subsection{Inverse problem for SSNP-sIDT}\label{sec:inv-sIDT}

Next, we formulate the inverse problem for reconstructing the 3D RI distribution using multiple intensity measurements taken from $L$ different LEDs $I_\mathrm{meas}^l$ indexed by $l=1, 2, \dots, L$.
We formulate the sequential IDT reconstruction as the following optimization problem
\begin{equation}\label{optim}
  \hat{n} = \underset{n\in \Theta}{\operatorname{argmin}}\left\{
    \sum_{l=1}^{L}\left\lVert \sqrt{\vIssnp^l} - \sqrt{I_\mathrm{meas}^l} \right\rVert_2^2
+\tau R_\mathrm{TV}(n)
\right\},
\end{equation}
where $\Theta$ is a set to enforce physical constraints, such as realness for non-absorbing samples and positivity for certain samples.
The first data-fidelity term computes the \mbox{$L^2$-norm} of the difference between the forward model computed and measured amplitude.
We use the magnitude instead of intensity based on a previous observation that it is more robust to signal dependent noise~\cite{Yeh:15}.
The second regularization term utilizes priors about the sample to alleviate the ill-posedness of the inverse problem.
Here we adopt the widely used 3D total variation regularizer $R_\mathrm{TV}$ that assumes the RI distribution is piece-wise constant~\cite{Kamilov2016,Chen:20,Chowdhury2019}.
One can also use more advanced ``denoisers'', such as denoising deep networks, to further boost the performance, as shown in recent works~\cite{Wu2020}. $\tau$ is a tuning parameter that controls the strength of the regularization.

We calculate the gradient using the automatic differentiation technique.
Since the SSNP forward model consists of a chain of simple operations, the gradient of the data-fidelity term can be computed using the chain rule to perform ``error back-propagation''.
We define the gradient $\Grad$ of a complex function $\cvary(\cvarx)$ as $\Grad(\cvary,\cvarx)=\overline{\partial \cvary / \partial \cvarx+\partial\overline{\cvary}/\partial\cvarx}$ according to the Wirtinger calculus~\cite{Wir1927}, where the $\overline{\cdot}$ operator means the complex conjugate.
The chain rule of the gradient is
\begin{equation} {\label{eq:chain}}
  \Grad(\cvary,\cvarz)=\Grad(\cvary,\cvarx)\overline{\diffp{\cvarx}{\cvarz}}.
\end{equation}
By repeatedly applying the chain rule in Eq.~\eqref{eq:chain}, one can compute the gradient from a single LED measurement.
For notation simplicity, we omit the LED index in the subsequent derivation.
Considering the data-fidelity contribution from a single LED\@:
$\Loss = \left\lVert \sqrt{\vIssnp}-\sqrt{I_\mathrm{meas}} \right\rVert_2^2$,
our goal is to get $\diffp{\Loss}/{{n(x,y,z_i)}}$, which is the gradient for the slice $z_i$.
This gradient can be computed from a series of local gradient computed using its definition and the chain rule, as follows:
\begin{align}
  &\Grad(\Loss, \vPhifzc)
  =2\overline{\diffp{\Loss}{\vPhifzc}}
  =2\left(\left|\vPhifzc\right|-\sqrt{I_\mathrm{meas}}\right)
  \frac{\vPhifzc}{\left|\vPhifzc\right|},
\end{align}
\begin{align}\label{core_start}
  &\Grad(\Loss,\vPhi_\mathrm{cam})
  =\overline{\diffp{\vPhifzc}{\vPhi_\mathrm{cam}}}\ \Grad(\Loss, \vPhifzc)
  =\iFxy{\begin{pmatrix}1/2\\j/(2k_z)\end{pmatrix}\Fxys{\Grad(\Loss, \vPhifzc)}},\\
  &\nonumber\Grad(\Loss,\vPhi_{xy}(z_n))
  =\overline{\diffp{\vPhi_\mathrm{cam}}{\vPhi_{xy}(z_n)}}\ \Grad(\Loss, \vPhi_\mathrm{cam})\\
  &=\iFxy{
    \begin{pmatrix}
      \cos(k_z \Delta z_f)&-k_z\sin(k_z \Delta z_f)\\
      \sin(k_z \Delta z_f)/k_z\!&\cos(k_z \Delta z_f)
  \end{pmatrix}\cdot\Fxys{\mathcal{P}_{\mathrm{NA}}\ \Grad(\Loss, \vPhi_\mathrm{cam})}}.
\end{align}
Next, we calculate a series of gradient $\Grad(\Loss,\vPhi_{xy}(z_i))$ indexed by $i={n-1}, \ldots, 2, 1 $ (corresponding to a physically reversed order), using the chain rule.
For notational simplicity, we define $\Grad_i\equiv\Grad(\Loss,\vPhi_{xy}(z_i))$ and $\Grad'_i\equiv\Grad(\Loss,\vQ(z_i)\vPhi_{xy}(z_i))$, and derive the following relations,
\begin{align}\label{G_i}
  \Grad'_i\nonumber
  &=\overline{\diffp{\vPhi_{xy}(z_{i+1})}{[\vQ(z_i)\vPhi_{xy}(z_i)]}}\ \Grad_{i+1}
  =\overline{\diffp{[\vP\,\vQ(z_i)\vPhi_{xy}(z_i)]}{[\vQ(z_i)\vPhi_{xy}(z_i)]}}\ \Grad_{i+1}\\
  &=\iFxy{\begin{pmatrix}
      \cos(k_z \Delta z)&-k_z\sin(k_z \Delta z)\\
      \sin(k_z \Delta z)/k_z\!&\cos(k_z \Delta z)
  \end{pmatrix}\Fxys{\Grad_{i+1}}},\\[1em]
  \Grad_i
  &=\overline{\diffp{[\vQ(z_i)\vPhi_{xy}(z_i)]}{\vPhi_{xy}(z_i)}}\ \Grad'_i
  =\begin{pmatrix}
    1&k_0^2\left(n_0^2-n_{xy}^2(z)\right)\Delta z\\
    0&1
  \end{pmatrix}\Grad'_i,
\end{align}
where $n_{xy}(z_i)\equiv n(x,y,z_i)$.
The gradient of the data-fidelity term from this measurement is
\begin{align}\label{ngrad}
  \Grad(\Loss,n_{xy}(z_i))
  &\nonumber=\overline{\diffp{\vQ(z_i)}{n_{xy}(z_i)}}
  \;\;\overline{\diffp{[\vQ(z_i)\vPhi_{xy}(z_i)]}{\vQ(z_i)}}\ \Grad'_i\\
  &\nonumber=\Re \left(-2\, k_0^2\, n_{xy}(z_i) \Delta z\ \overline{\vPhi_{xy}(z_i)}^\top
    \begin{pmatrix}0&1\\0&0\end{pmatrix} \Grad'_i
  \right)\\
  &=\Re \left(-2\,k_0^2\,n_{xy}(z_i)\Delta z\ \left(0\quad\varphi_{xy}(z_i)\right) \Grad'_i\right),
\end{align}
where ${(\cdot)}^\top$ denotes matrix transpose.
Finally, the total gradient of the entire data-fidelity term is simply the sum over all the measurements. A diagram that illustrates the forward and gradient calculation is provided in Supplement 1.

With the gradient of the data-fidelity term, we perform the reconstruction using the fast iterative shrinkage-thresholding algorithm (FISTA) to solve Eq.~\ref{optim}.
The 3D TV regularization is implemented using an efficient proximal operator based on~\cite{Kamilov2016}.
We initialize $n(\bm{r})$ as the background RI, $n_0$, and then update the estimation iteratively.

\subsection{Inverse problem for SSNP-mIDT}

For multiplexed IDT data, we have intensity measurements from $L$ illumination patterns, with each pattern containing $M$ LEDs.
Correspondingly, the reconstruction is achieved by solving the following optimization problem:
\begin{equation}\label{optim_midt}
  \hat{n} = \underset{n\in \Theta}{\operatorname{argmin}}\left\{
    \sum_{l=1}^{L}\left\lVert \sqrt{\sum_{m=1}^M \vIssnp^{l,m}} - \sqrt{I_\mathrm{meas}^l} \right\rVert_2^2
+\tau R_\mathrm{TV}(n)
\right\},
\end{equation}
where $I_\mathrm{meas}^l$ denotes the $l$th intensity measurement and $\vIssnp^{l,m}$ denotes the SSNP-model computed intensity from the $m$th LED in the $l$th pattern.

To derive the gradient for the multiplexed data-fidelity term, we discuss a single LED pattern and omit the pattern index in the subsequent derivation for brevity.
The data-fidelity term for the $l$th measurement is
$\Loss' = \left\lVert \sqrt{\sum_{m=1}^M \vIssnp^m}-\sqrt{I_\mathrm{meas}} \right\rVert_2^2$,
and the local gradient is
\begin{align}
  \Grad(\Loss', \vPhifzc^m)
  &=2\ \overline{\diffp{\Loss'}{\vPhifzc^m}}
  =2\ \overline{\diffp{\Loss'}{\vIssnp^m}}\cdot\overline{\diffp{\vIssnp^m}{\vPhifzc^m}} \nonumber
  \\
  &=2\ \frac{\sqrt{\sum_{m=1}^M \vIssnp^m}-\sqrt{I_\mathrm{meas}}}{\sqrt{\sum_{m=1}^M \vIssnp^m}}\cdot\vPhifzc^m.
\end{align}
This gradient step effectively performs demultiplexing, which is analogous to the 2D multiplexed phase retrieval problem~\cite{Tian:14}.
The subsequent gradient calculation steps are the same as Eq.~\eqref{core_start}-\eqref{ngrad} except that $\vPhifzc$ is replaced by $\vPhifzc^m$ and the gradient $\Grad(\Loss',n_{xy}(z_i))$ calculated from all the LEDs in each measurement is summed up at the final step.

Overall, our SSNP-IDT algorithm is computationally efficient and requires $\sim2\times$ time and memory cost as the BPM algorithm since the SSNP requires computing both the field and its axial derivative.
We implemented the algorithm using custom CUDA code on Python to leverage the massively parallel processing power of GPU\@.
To reconstruct a single volume containing $\num{600 x 600 x 150}$ voxels, the algorithm typically requires 20 iterations to converge and takes $\sim \SI{120}{\s}$ on a PC equipped with a GeForce RTX2070 GPU\@.
A unified efficient algorithm for both the sequential and multiplexed IDT and an evaluation of the computational performance are provided in Supplement 1.

\section{Results}
\subsection{Simulation: forward model accuracy evaluation}

\begin{figure}[t]
  \centering
  \includegraphics[width=0.8\textwidth]{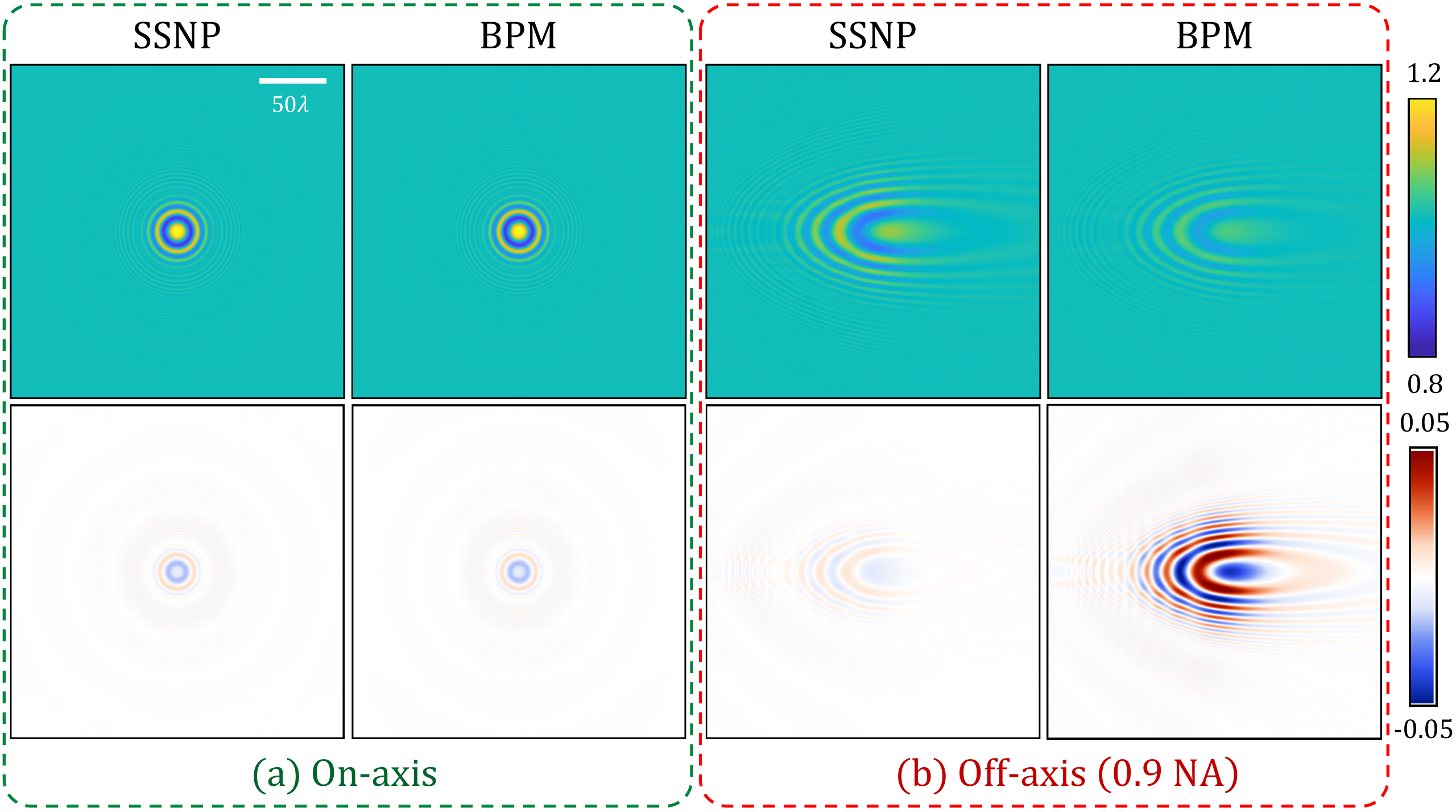}
  \caption{
    Simulation using the SSNP and BPM IDT models for a $6\lambda$ diameter bead ($n=1.02$) in air ($n_0=1$) using (a) on-axis illumination (NA = 0), and (b) off-axis illumination (NA = 0.9).
    The bead center is placed at the origin at $z=0$, and the observation plane is placed at $z=125\lambda$.
    The upper row shows the normalized intensity, and the lower row shows the difference comparing against the Mie theory.
  }\label{figm}
\end{figure}

A major benefit of the SSNP model compared with the BPM model is its higher accuracy when computing scattered field under high-NA illumination~\cite{Lim2019}.
To evaluate the SSNP-IDT model's accuracy under high-NA illumination, we simulated sequential IDT measurements of a bead using the BPM and SSNP models and compare them against the Mie scattering theory.
In this simulation, the bead has an RI of 1.02 and is immersed in air ($n_0=1$).
The diameter of the bead is $6\lambda=\SI{3.09}{\um}$ where $\lambda=\SI{0.515}{\um}$ is the wavelength.
In Fig.~\ref{figm}, we show the on-axis and an off-axis 0.9~NA illumination case.
The microscope has a 0.9~NA objective lens.
The discretization steps are $(\Delta x, \Delta y, \Delta z)=(\lambda/4, \lambda/4, \lambda/8)$.
The simulated intensities from the BPM and SSNP IDT models and their difference maps from the Mie theory are shown.
For the on-axis illumination case, both the SSNP and BPM models agree with the Mie theory with negligible errors.
For the off-axis high-NA illumination case, SSNP retains high accuracy while the BPM model suffers from more severe errors.
This study confirms that our SSNP-IDT forward model is highly accurate in simulating IDT measurements under high-NA illuminations.
As a result, we will use the SSNP model for simulating the measurements in the subsequent numerical studies due to its higher computational efficiency and flexibility compared to the Mie theory.


\subsection{Simulation: reconstruction algorithm evaluation}

\begin{figure}[htb]
  \centering
  \includegraphics[width=0.9\textwidth]{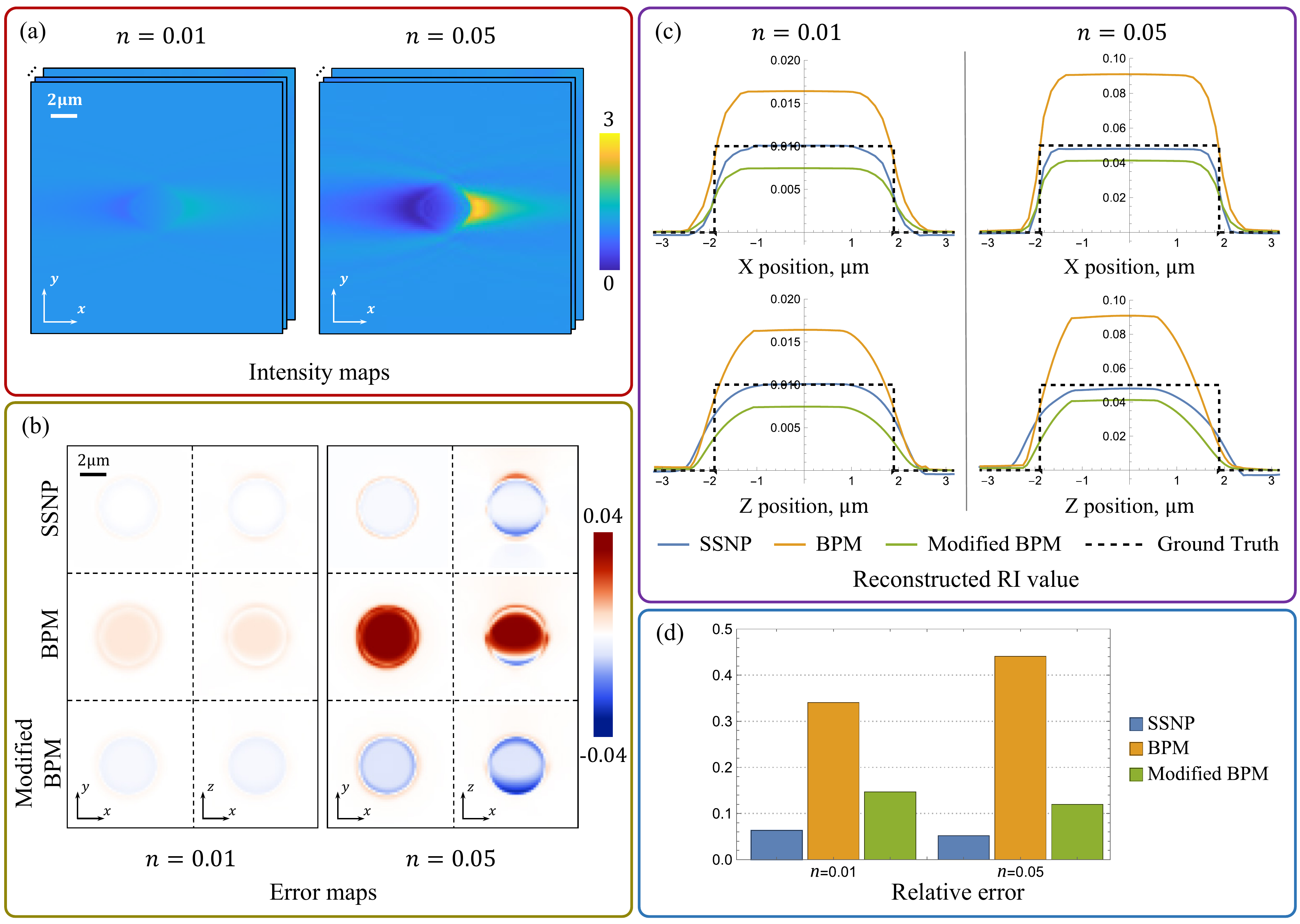}
  \caption{
    (a) An example image from the SSNP simulated intensity measurements for low and high RI sphere samples.
    (b) Error maps of the RI reconstructions from the SSNP, original BPM, and modified BPM models.
      The X-Y and X-Z cross sections through the center of each sphere are shown.
    (c) The cutlines along X and Z of the reconstructed and ground-truth RI through the center of each sphere.
    (d) The relative MSE of all the reconstructions.
  }\label{figsim}
\end{figure}

To assess our SSNP-IDT reconstruction algorithm in the non-paraxial regime, we simulated high-NA annular IDT measurements using the SSNP-sIDT forward model and then performed reconstruction using the algorithm in Section~\ref{sec:inv-sIDT}.
We simulated a $6\lambda$-diameter sphere with voxel size $(\lambda/4, \lambda/4, \lambda/8)$.
To study the performance under different scattering strengths, we used two RI contrast values, including $\Delta n$=0.01 and 0.05.
We simulated 8 intensity images from illuminations evenly distributed on a 0.89 NA ring, collected by a 0.9 NA objective (Fig.~\ref{figsim}a).

We performed reconstruction using both the SSNP, the original BPM~\cite{Kamilov2016}, and modified BPM model~\cite{Lim2019} (see details in Supplement 1) in Fig.~\ref{figsim}b and~\ref{figsim}c.
Then we use the relative mean squared error (MSE) $\mathscr{E}$ to quantify the overall reconstruction quality in Fig.~\ref{figsim}d, which is defined as
$
  \mathscr{E}=\frac{{\left\lVert{n_{GT}-\hat{n}}\right\rVert}^2}{{\left\lVert{n_{GT}-n_0}\right\rVert}^2},
$
where $n_{GT}$ and $\hat{n}$ are the ground-truth and reconstructed RI, respectively.
The original BPM over-estimates the RI values in both the weakly and strongly scattering cases.
This is because the BPM forward model underestimates the phase modulation for the high-NA components~\cite{Lim2019}.
The modified BPM model introduces an cosine obliquity factor to compensate for the under-estimated phase modulation, which improves the model accuracy at high angles~\cite{Lim2019}.
However, this approximation results in under-estimation in the reconstructed RI values.
Figure~\ref{figsim}d quantitatively shows that the SSNP model can recover the RI distribution with better accuracy compared to the BPM and modified BPM models in both the weakly and strongly scattering cases.
By inspecting the cross sections and cutlines shown in Fig.~\ref{figsim}b and~\ref{figsim}c, we concluded that the reconstruction in all three dimensions are limited by the finite Fourier coverage provided by the illumination and imaging optics.
Both the lateral resolution and the amount of axial elongation due to the ``missing-cone'' problem are comparable regardless of the model used.

\subsection{Experiments: sequential IDT}
To test our SSNP-sIDT reconstruction algorithm, we apply it to the annular IDT data from our previous publication in~\cite{Li2019}.
Briefly, our annular IDT system consists of a Nikon E200 microscope equipped with a ring LED (Adafruit, 1586 NeoPixel Ring).
We used a $40\times$/$0.65$ NA (Nikon, CFI Plan Achromat) objective.
The ring LED has 24 LEDs and is $\SI{60}{\mm}$ in diameter.
It is centered at the optical axis and placed approximately $\SI{35}{\mm}$ away from the sample, which sets the illumination angle to be $\sim \ang{40}$ and matches with the objective NA\@.
Each LED illumination is approximated as a plane wave with a wavelength $\lambda=\SI{515}{\nm}$.
We use the same self-calibration method as \cite{Li2019} to get the accurate direction of the wave vector.

In our reconstruction algorithm, the physical quantities on the XY plane, including the field, the field's axial derivative, and the slices of the predicted RI values, are discretized on 2D grids.
The grid spacing is $\Delta x=\Delta y=\SI{0.1625}{\um}$, which is chosen to be smaller than but close to $\frac{\lambda}{4 \mathrm{NA}}$ to ensure both accuracy and computational efficiency of the algorithm, and is the same as the effective pixel size on the camera.

\begin{figure}[htb]
  \centering
  \includegraphics[width=\textwidth]{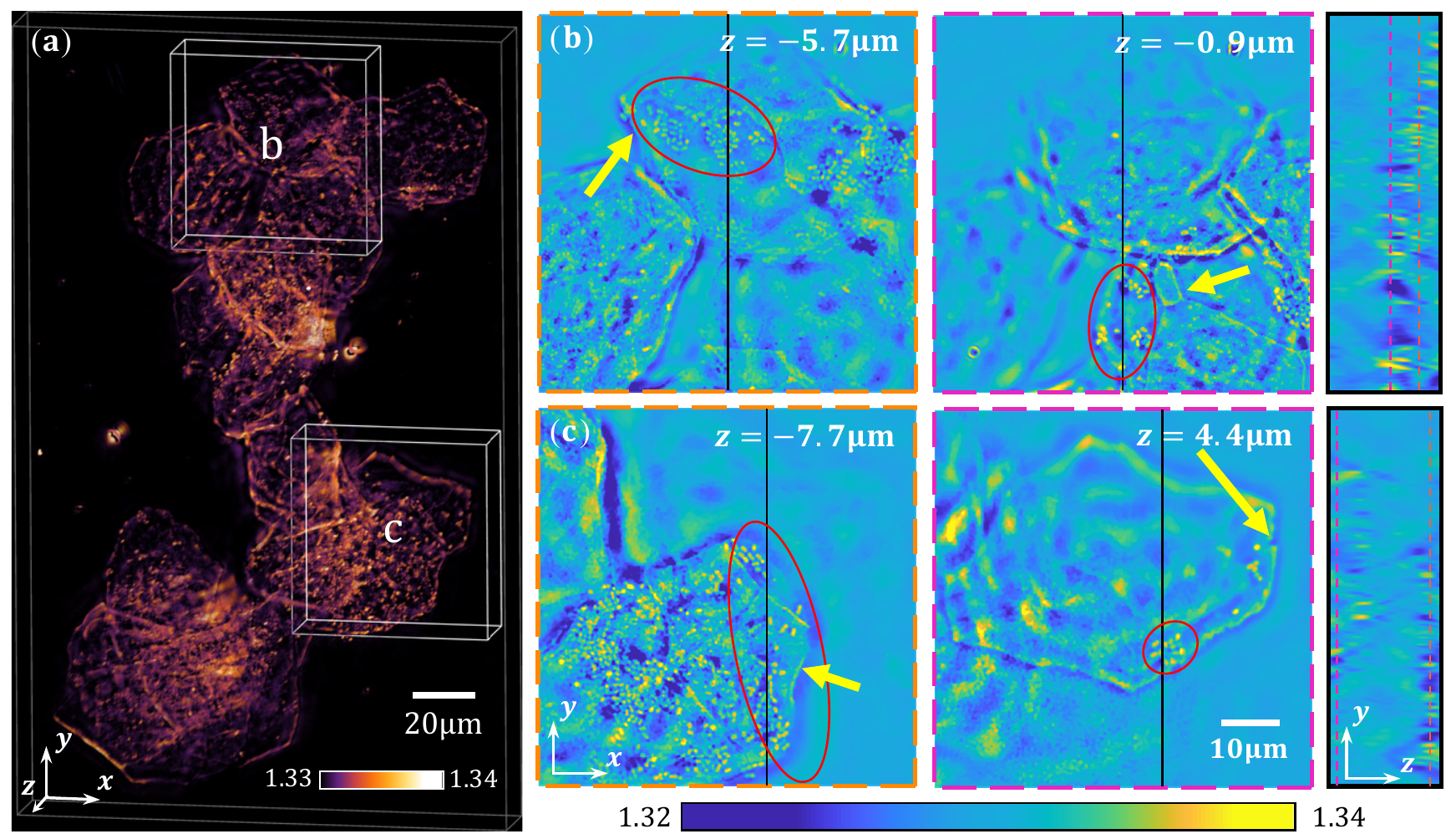}
  \caption{Reconstruction of buccal epithelial cells.
  (a) 3D rendering of the full-FOV reconstruction.
  (b), (c) Zoomed-in views of XY (Top), XZ (Bottom) cross sections. The red ellipses show the native bacteria crowded regions. The yellow arrows highlight the sharp cell boundary features.
}\label{fig_a}
\end{figure}

First, we verify the reconstruction algorithm on unstained buccal epithelial cells immersed in purified water on a glass slide and covered by a coverslip.
The dataset contains $24$ intensity images.
The reconstructed volume is between $\SIrange{-9.7}{9.7}{\um}$ around the focal plane, and the FOV is $\SI{162.5 x 162.5}{\um}$.
We discretized the sample volume to $150$ slices, each with $1000 \times 1000$ pixels, with voxel size $\SI{0.1625 x 0.1625 x 0.129}{\um}$.
To perform the reconstruction with memory-limited hardware, we cropped the full FOV into four $\num{576 x 576}$-pixel subregions, reconstructed them separately, and then stitched them together.
Figure~\ref{fig_a}(a) shows the 3D rendering of the reconstructed RI distribution of the entire cell cluster volume.
The 3D reconstruction allows easily discriminating cells at different depths.
As shown in Fig.~\ref{fig_a}(b-c), our SSNP algorithm successfully reconstructs high-resolution features, such as cell boundaries, membrane, and native bacteria around the cells.

\begin{figure}[ht!]
  \centering
  \includegraphics[width=\textwidth]{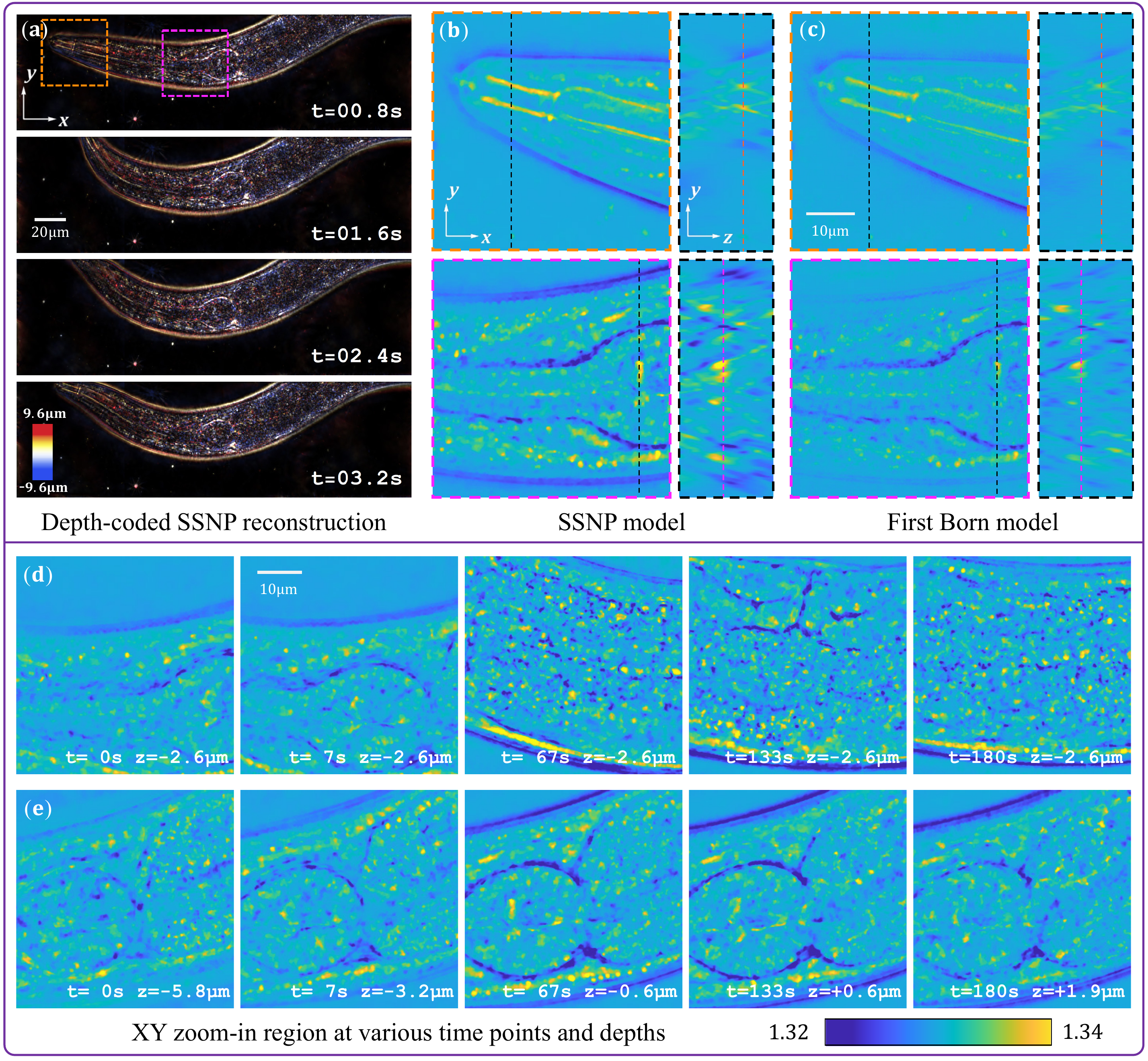}
  \caption{Reconstruction of a live \textit{C. elegans} worm.
  (a) Color-coded depth projections of the whole worm reconstruction using SSNP model at different times.
  (b-c) XY and ZY cross-sections of the buccal cavity (top), part of the isthmus and terminal bulb (bottom).
    The reconstructions are using SSNP model and first-Born based model, respectively.
  (d) XY cross-sections of SSNP-IDT reconstruction at different time points of the same subregion and depth.
  (e) XY cross-sections of SSNP-IDT reconstruction at different depths and time points of the same subregion.
}\label{fig4}
\end{figure}

Next, we apply our algorithm to a live \textit{C. elegans} worm sample.
Young adult \textit{C. elegans} were mounted on $\SI{3}{\percent}$ agarose pads in a drop of nematode growth medium (NGM) buffer.
Glass coverslips were then gently placed on top of the pads and sealed with a 1:1 mixture of paraffin and petroleum jelly.
The time series IDT measurements were taken at $\sim \SI{10}{\hertz}$ with 8 intensity images for each set.
The reconstruction is performed in a volume of $\SI{97.5 x 390 x 38.6}{\um}$ centered at the focal plane.
The volume is discretized to $\num{600 x 2400 x 300}$ voxels with voxel size $\SI{0.1625 x 0.1625 x 0.129}{\um}$.
The reconstruction is more challenging since the worm contains complex and strongly scattering features.
Figure~\ref{fig4}(a) shows four consecutive volumes reconstructed by our SSNP algorithm, displayed as color-coded depth projections.
Fig.~\ref{fig4}(b-c) shows two zoomed-in XY and YZ cross-sectional views around the buccal cavity and terminal pharyngeal bulb regions.
Comparing with the SSNP model, the first-Born model underestimates the RI value and misses the low-frequency information.
This effect is also reported in other multiple-scattering models for IDT reconstruction~\cite{Chowdhury2019, Chen:20}.
Additional zoom-in regions at different depths and time points are shown in Fig.~\ref{fig4}(d-e), containing features including the anterior bulb, vulva, lipid droplets, and body wall muscle cells inside the worm sample.

\subsection{Experiment: multiplexed IDT}
Next, we test our SSNP-mIDT reconstruction algorithm to process data from our previous publication in~\cite{Matlock2019}.
Briefly, the multiplexed IDT system consists of a Nikon TE 2000U microscope equipped with a custom LED matrix.
Each LED approximately generated a plane wave with a central wavelength $\lambda=\SI{632}{\nm}$.
The measurements were taken with a 40$\times$~/~0.65~NA objective (Nikon, CFI Plan Achromat) and an sCMOS camera (PCO.Edge 5.5).

\begin{figure}[b]
  \centering
  \includegraphics[width=0.9\textwidth]{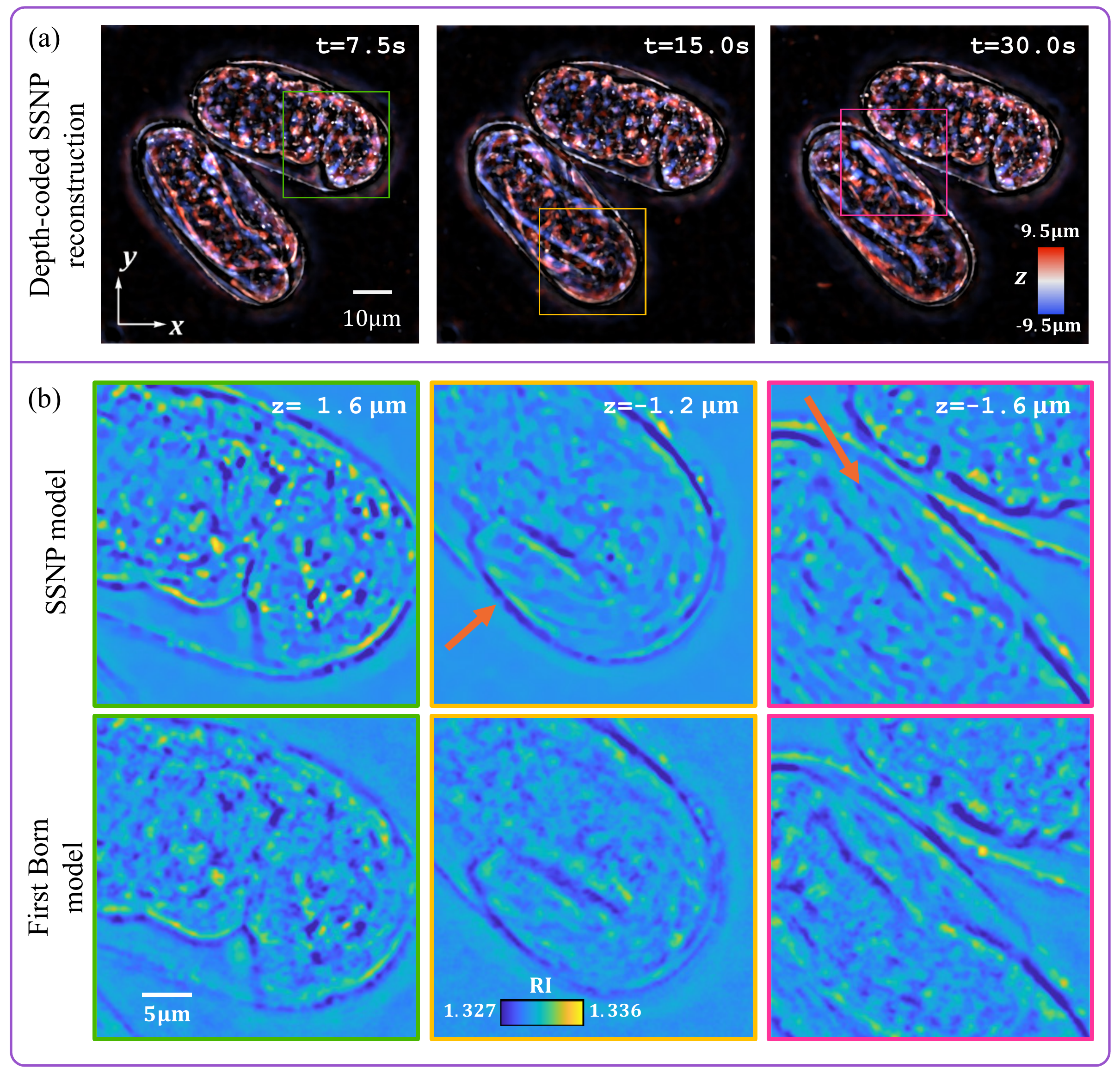}
  \caption{Reconstruction of live \textit{C. elegans} embryos.
    (a) Color-coded depth projection of SSNP-mIDT reconstructon at $t=\SIlist{7.5;15;30}{\s}$
    (b-c) XY zoom-in cross sections at various times and depths using SSNP and first-Born models, which shows developing cells, buccal cavity and tail region of the embryos and worms, respectively.
}\label{fig5}
\end{figure}

We apply our algorithm to a highly scattering sample of live \textit{C. elegans} embryos.
The multiplexed measurement used $16$ disjoint illumination patterns, each containing 6 LEDs.
In total 96 LEDs are used in this experiment, corresponding to illumination NA ranging from $0.3$ to $0.575$.
We reconstructed a $\SI{81.3 x 81.3 x 19.0}{\um}$ volume, which was discretized to $\num{500 x 500 x 120}$ voxels with voxel size $\SI{0.1625 x 0.1625 x 0.158}{\um}$.
In Fig.~\ref{fig5}, the embryo on the left is in the late three-fold (quickening) stage, and the one on the right is in the comma stage.
From the color-coded view in Fig.~\ref{fig5}(a), how the worms are folded can be clearly observed.
From the single-depth cross-sections in Fig.~\ref{fig5}(b), the morphological details of the cells' outline, the buccal cavity, and the tail of the worm are reconstructed.
Similar to aIDT, the SSNP model reconstructs more low-frequency phase information and higher RI contrast than the first-Born model.
Besides, the TV regularization used in the SSNP model effectively suppresses the noise, which helps to provide a clean background and sharp embryo boundary comparing with the Tikhonov regularization used in the first-Born model.
The high acquisition rate of multiplexed IDT enabled reconstructing the developing process and the moving behavior of the embryos, as shown in the time-series reconstructions.
As compared to our previous single-scattering based linear multiplexed IDT reconstruction algorithm~\cite{Matlock2019}, SSNP-mIDT provides more robust demultiplexing capability that removes the fuzzy diffraction artifacts inside the embryos.
Since the algorithm allows independent update from individual LEDs in each pattern, it greatly suppresses the cross-talk artifacts suffered by the linear model.

\section{Conclusion and discussion}

In conclusion, we developed a non-paraxial multiple-scattering model for 3D RI reconstruction from both sequential and multiplexed intensity diffraction tomography measurements.
The SSNP model ensures high accuracy when imaging strongly scattering samples under high-angle illumination.
We demonstrated the robustness of the model on weakly scattering cells and multiple-scattering dynamic \textit{C. elegans} worm and embryos.
With a unified reconstruction algorithm, our model is flexible for both the sequential and multiplexed IDT setups.
This treatment for multiplexed measurement can also be applied to interferometric setups and speed up the acquisition for the ODT systems~\cite{Park2018, Lim2019}.

A major limitation of our current IDT systems is the limited angular coverage (up to 0.65 NA) that creates a significant amount of missing-cone artifacts and fundamentally limits the reconstruction quality, as shown by our numerical studies and experimental results.
While several recent works alleviate the missing-cone artifacts by using higher NA (>1) optics and more than 10$\times$ more measurements~\cite{Chowdhury2019,Chen:20}, it sacrifices the FOV and acquisition speed, which prevented them from capturing the whole worm without mechanical FOV scanning and imaging dynamic biological processes.
It may be possible to integrate the annular and multiplexed illumination strategies in these systems to speed up the acquisition while achieving better Fourier coverage.

It has been shown that 3D Fourier Ptychography models based on the first-Born approximation~\cite{ZUO2020106003} and BPM~\cite{Tian2015} can further handle dark-field measurements, which allows considering high spatial frequency information beyond the objective NA in the 3D reconstruction.
Since our SSNP model can be treated as an enhancement of the first-Born and BPM models, it is possible to extend our work to incorporate additional dark-field information for high-resolution IDT reconstructions, which will be considered in our future work.
Alternatively, several advanced reconstruction frameworks for IDT have been recently developed and shown promising results to greatly suppress the missing-cone artifacts, including end-to-end supervised learning~\cite{matlock21}, untrained neural network~\cite{Zhou:20}, model-based reconstruction with a deep denoiser~\cite{simba}, neural fields based reconstruction~\cite{decaf21}, and physics-informed neural network~\cite{ayoub_optical_nodate}.
Notably, our SSNP forward model has been previously applied to generate a large-scale training data set to train a supervised learning model~\cite{matlock21}. It is further possible to directly incorporate our SSNP model into the model-based learning strategies~\cite{Zhou:20,simba,decaf21} in the future.

\begin{backmatter}
\bmsection{Funding}
National Science Foundation: 1846784.


\bmsection{Disclosures}
The authors declare no conflicts of interest.

\bmsection{Data availability}
The data and code can be found at the GitHub repository:
\href{https://github.com/bu-cisl/SSNP-IDT}{https://github.com/bu-cisl/SSNP-IDT}

\bmsection{Supplemental document}
See Supplement 1 for supporting content.

\end{backmatter}

\bibliography{reference}

\end{document}


\maketitle

\section{Details of SSNP model}
We rewrite the Helmholtz equation into the matrix form
\begin{equation}\label{sup:diff}
  \diffp{\vPhi(\bm{r})}z=\bm{\mathrm{H}}(\bm{r})\;\vPhi(\bm{r}),
\end{equation}
where
\begin{equation}\label{sup:SSNP2}
\vPhi(\bm{r})=\begin{pmatrix}\varphi(\bm{r})\\ \frac{\partial \varphi}{\partial z}\end{pmatrix},
\ \vH(\bm{r})=\begin{pmatrix}
  0&1\\-\diffp[2]{}x-\diffp[2]{}y-\,k_0^2  n^2(\bm{r})\!\!&0
  \end{pmatrix}.
\end{equation}
This is a first-order homogeneous linear system of differential equations with variable coefficients for the variable $z$.
We assume $n(\bm{r})$ is continuous.
For a sufficiently small $\Delta z$, we can get the following stepwise solution:
\begin{equation}
  \vPhi_{xy}(z_i+\Delta z)=\exp[\int_{z_i}^{z_i+\Delta z} \vH(\bm{r}) \,dz]\,\vPhi_{xy}(z_i) + O(\Delta z)=\exp(\vH(\bm{r})\Delta z)\,\vPhi_{xy}(z) + O(\Delta z).
\end{equation}
To compute the traverse Laplace operator $\diffp[2]{}x+\diffp[2]{}y$ efficiently, we need to transform it into the Fourier space while leaving the other variable part in the image space.
Thus, we decouple $\exp(\vH)$ into $\exp(\vH_1)$ and $\exp(\vH_2(\bm{r}))$ where
\begin{equation}\label{sup:decop}
  \vH_1=
  \begin{pmatrix}
    0&1\\-\diffp[2]{}x-\diffp[2]{}y\!-\!\,k_0^2 n_0^2\!&0
  \end{pmatrix},\,
  \vH_2(\bm{r})=
  \begin{pmatrix}
    0&0\\k_0^2(n_0^2-n^2(\bm{r}))\!&0
  \end{pmatrix}.
\end{equation}
Since $\vH_1$ and $\vH_2$ do not commute, the decoupling only holds approximately when $\Delta z$ is small:
\begin{equation}
  \exp(\vH(\bm{r})\Delta z)\,\vPhi_{xy}(z)=\vP\,\vQ(\bm{r})\,\vPhi_{xy}(z)+O([\vH_1, \vH_2] {\Delta z}^2),
\end{equation}
where $\vP = \exp(\vH_1\Delta z),\ \vQ(z) = \exp(\vH_2(x,y,z)\Delta z)$, and $[\vH_1, \vH_2] = \vH_1 \vH_2 - \vH_2 \vH_1$ is the commutator for matrix $\vH_1$ and $\vH_2$.
This also suggest that one should choose the constant $n_0$ in~\eqref{sup:decop} as the background or average RI of the media to keep $[\vH_1, \vH_2]$ as small as possible.
After we decoupled $\exp(\vH)$, we can compute the operator $\vP$ and $\vQ$ separately.
In the Fourier space, the operator $\widetilde{\vH_1}$ is a matrix function with frequency component $(k_x, k_y)$:
\begin{equation}
  \widetilde{\vH_1}=
  \begin{pmatrix}
    0&1\\k_x^2+k_y^2-k_0^2 n_0^2\!&0
  \end{pmatrix}.
\end{equation}
We diagonize $\widetilde{\vH_1}$ to compute the matrix exponential:
\begin{align} \label{eq:tildeP}
  \nonumber\widetilde{\vP}=\exp(\widetilde{\vH_1} \Delta z) &= \exp\left[
    \begin{pmatrix}1 & 1 \\ j\,k_z & -j\,k_z\end{pmatrix}
    \begin{pmatrix}j\,k_z\,\Delta z & 0 \\ 0 & -j\,k_z\,\Delta z\end{pmatrix}
    \begin{pmatrix}1/2 & -j/(2\,k_z) \\ 1/2 & j/(2\,k_z)\end{pmatrix}
  \right]\\
  \nonumber
  &=\begin{pmatrix}1 & 1 \\ j\,k_z & -j\,k_z\end{pmatrix}
    \begin{pmatrix}\exp(j\,k_z\,\Delta z) & 0 \\ 0 & \exp(-j\,k_z\,\Delta z)\end{pmatrix}
    \begin{pmatrix}1/2 & -j/(2\,k_z) \\ 1/2 & j/(2\,k_z)\end{pmatrix}\\
  &=\begin{pmatrix}
      \cos(k_z \Delta z)&\sin(k_z \Delta z)/k_z\\
      -k_z\sin(k_z \Delta z)\!&\cos(k_z \Delta z)
    \end{pmatrix},
\end{align}
where $k_z=\sqrt{k_0^2\,n_0^2-k_x^2-k_y^2}$.
Thus, we have
\begin{equation}
  \vP\,\vPhi_{xy}=\iFxy{
    \begin{pmatrix}
      \cos(k_z \Delta z)&\sin(k_z \Delta z)/k_z\\
      -k_z\sin(k_z \Delta z)\!&\cos(k_z \Delta z)
    \end{pmatrix}
    \Fxys{\vPhi_{xy}}
  },
\end{equation}
where $\Fxys{\cdot}$ and $\iFxys{\cdot}$ denote the 2D Fourier and inverse Fourier transform on the XY plane, respectively.
We also define $\widetilde{\vP}=0$ for $k_x^2+k_y^2>k_0^2\,n_0^2$, which removes the evanescent component.

The high order matrix power of $\vH_2$ is $\bm{0}$ (i.e. $(\vH_2)^m$ = $\bm{0}$, for $m\ge2$), and we can use Taylor expansion to compute $\exp(\vH_2 \Delta z)$:
\begin{equation}
  \vQ = \mathbf{I} + \vH_2 \Delta z = \begin{pmatrix}1&0\\k_0^2\left(n_0^2-n_{xy}^2(z)\right)\Delta z&1\end{pmatrix}.
\end{equation}

\section{BPM and modified BPM}
The BPM algorithm also contains two computational steps in each slice. The first represents the propagation in the homogeneous background media and the other represents the local phase modulation due to the RI changes.
We use $\bpmP$ and $\bpmQ$ to denote the propagation and phase modulation operator, respectively.
The propagation step is performed by the angular spectrum method:
\begin{equation}\label{eq:bpmp}
  \psi_{xy}(z+\Delta z)=\bpmP \varphi_{xy}(z)=\iFxys{\exp\left(j\,\sqrt{k_0^2\,n_0^2-k_x^2-k_y^2}\Delta z\right) \varphi_{xy}(z)}.
\end{equation}
For the original BPM algorithm, we use the paraxial approximation to calculate the phase modulation:
\begin{equation}\label{eq:normbpm}
  \varphi_{xy}(z+\Delta z) = \bpmQ \psi_{xy}(z+\Delta z)=\exp\left[j\,k_0\,(n(\bm{r})-n_0)\Delta z\right]\psi_{xy}(z+\Delta z).
\end{equation}
This model assumes that the phase modulation caused by the RI change is invariant for different illumination angles. However, the phase change should vary with illumination angles. 
We can analyze this effect by introducing a small RI change in \eqref{eq:bpmp} by changing $n_0$ to $n_0 + \Delta n$.
The propagation operator in the Fourier space $\widetilde{\bpmP}$ becomes
\begin{align}\label{eq:mod1}
  \nonumber\widetilde{\bpmP'}&=\exp\left[j\,\sqrt{k_0^2\,(n_0+\Delta n)^2-k_x^2-k_y^2}\Delta z\right]\\
  \nonumber&\approx\exp\left[j(\sqrt{k_0^2\,n_0^2-k_x^2-k_y^2}+k_0^2\,n_0\,\Delta n/k_z)\Delta z\right]\\
  &=\widetilde{\bpmP}\exp(k_0 \Delta n \Delta z \cdot k_0\,n_0/k_z).
\end{align}
Comparing with \eqref{eq:normbpm}, there is an additional factor $k_0\,n_0/k_z=1/\cos \theta$ in the phase change, where $\theta$ is the angle between the wave vector $\bm{k}$ and the optical axis.
This obliquity factor captures the extra phase change because of the longer optical path as an oblique beam traveling through the sample slice.
The phase accumulation operator is more efficiently computed in the real space (Eq.~\ref{eq:normbpm}), but the obliquity factor in \eqref{eq:mod1} can only be calculated accurately in the Fourier space for all possible angles.
In weakly scattering case, most of the scattered energy stays around the illumination direction, so we can approximate the scattered beam propagation direction as the illumination direction and use a single obliquity factor $\cos \theta$.
Accordingly, the modified phase modulation operator $\mbpmQ$ is
\begin{equation}
  \mbpmQ(z)\psi_{xy}(z+\Delta z)=\exp\left[j\,k_0\,(n(\bm{r})-n_0)\Delta z / \cos \theta\right]\psi_{xy}(z+\Delta z).
\end{equation}
In summary, the slice-wise calculation used in the original BPM is:
\begin{equation}
  \varphi_{xy}(z+\Delta z)=\bpmP\,\bpmQ(z)\,\varphi_{xy}(z),
\end{equation}
and the modified BPM is:
\begin{equation}
  \varphi_{xy}(z+\Delta z)=\bpmP\,\mbpmQ(z)\,\varphi_{xy}(z).
\end{equation}

\section{Connection between z-derivation and bi-directional propagation in SSNP}
The axial derivative of the field can be represented by the forward-propagating and back-propagating components of the field.
Considering the field of a single plane wave $\varphi(\bm{r};k_x,k_y)$
\begin{align}
  &\varphi_{plane}(r)=\varphi_0 \exp(j(x k_x + y k_y \pm z \sqrt{k_0^2\,n_0^2 - k_x^2 - k_y^2})),\\
  \label{dpdz}&\diffp{\varphi_{plane}(r)}{z}= \pm j\sqrt{k_0^2\,n_0^2 - k_x^2 - k_y^2} \varphi_{plane}(r).
\end{align}
For the positive square-root case, the direction of $\bm{k}$ vector is in the $+z$ half-space, which means that the wave is propagating forward.
Otherwise, it means the wave is propagating backward.
According to angular spectrum theory, an arbitrary field can be represented by the sum of many plane wave components.
We can obtain the following relation by invoking linearity and \eqref{dpdz}:
\begin{align}\label{eq:fb_decompose}
  \Fxys{\vPhi_{xy}(z)}=\begin{pmatrix}\Fxys{\varphi}\\\Fxys{\diffp{\varphi}{z}}\end{pmatrix}=
  \begin{pmatrix}1&1\\j k_z & -j k_z\end{pmatrix}
    \begin{pmatrix}\Fxys{\varphi_f}\\\Fxys{\varphi_b}\end{pmatrix}.
\end{align}
As a special case, for any forward propagating illumination field $\varphi_{xy}^{in}$, it can be converted to
\begin{equation}
  \vPhi_{xy} = \begin{pmatrix}\varphi_{xy}^{in} \\ \iFxys{j\,k_z\,\Fxys{\varphi_{xy}^{in}}}\end{pmatrix}.
\end{equation}
Without considering the evanescent wave, the transform in \eqref{eq:fb_decompose} is invertible:
\begin{align}\label{split_prop}
  \begin{pmatrix}\Fxys{\varphi_f}\\\Fxys{\varphi_b}\end{pmatrix}=
  \frac{1}{2} \begin{pmatrix}1 & -j/k_z\\1 & j/k_z\end{pmatrix}
  \begin{pmatrix}\Fxys{\varphi}\\\Fxys{\diffp{\varphi}{z}}\end{pmatrix} 
  = \frac{1}{2} \begin{pmatrix}1 & -j/k_z\\1 & j/k_z\end{pmatrix} \Fxys{\vPhi_{xy}(z)}.
\end{align}

One can relate the forward and backward propagating components to the propagation operator $\widetilde{\vP}$ of the SSNP by comparing \eqref{eq:fb_decompose} and \eqref{split_prop} with \eqref{eq:tildeP} in particular the second equality that performs eigenvalue decomposition.
One can conclude that $\varphi_f$ and $\varphi_b$ are the projection of $\vPhi_{xy}$ onto the two eigenvectors of $\widetilde{\vP}$ in the Fourier space.
Notably, the two eigenvalues of $\widetilde{\vP}$ are $\exp(-j\,k_z\,\Delta z)$ and $\exp(j\,k_z\,\Delta z)$, corresponding to the forward and backward propagating operator, respectively. 
This indicates that for the propagation operator, the SSNP is identical to computing both the forward and backward propagation using the BPM\@.

\section{Memory-Efficient modular automatic differentiation framework}

\begin{figure}[t]
  \centering
  \includegraphics[width=0.9\textwidth]{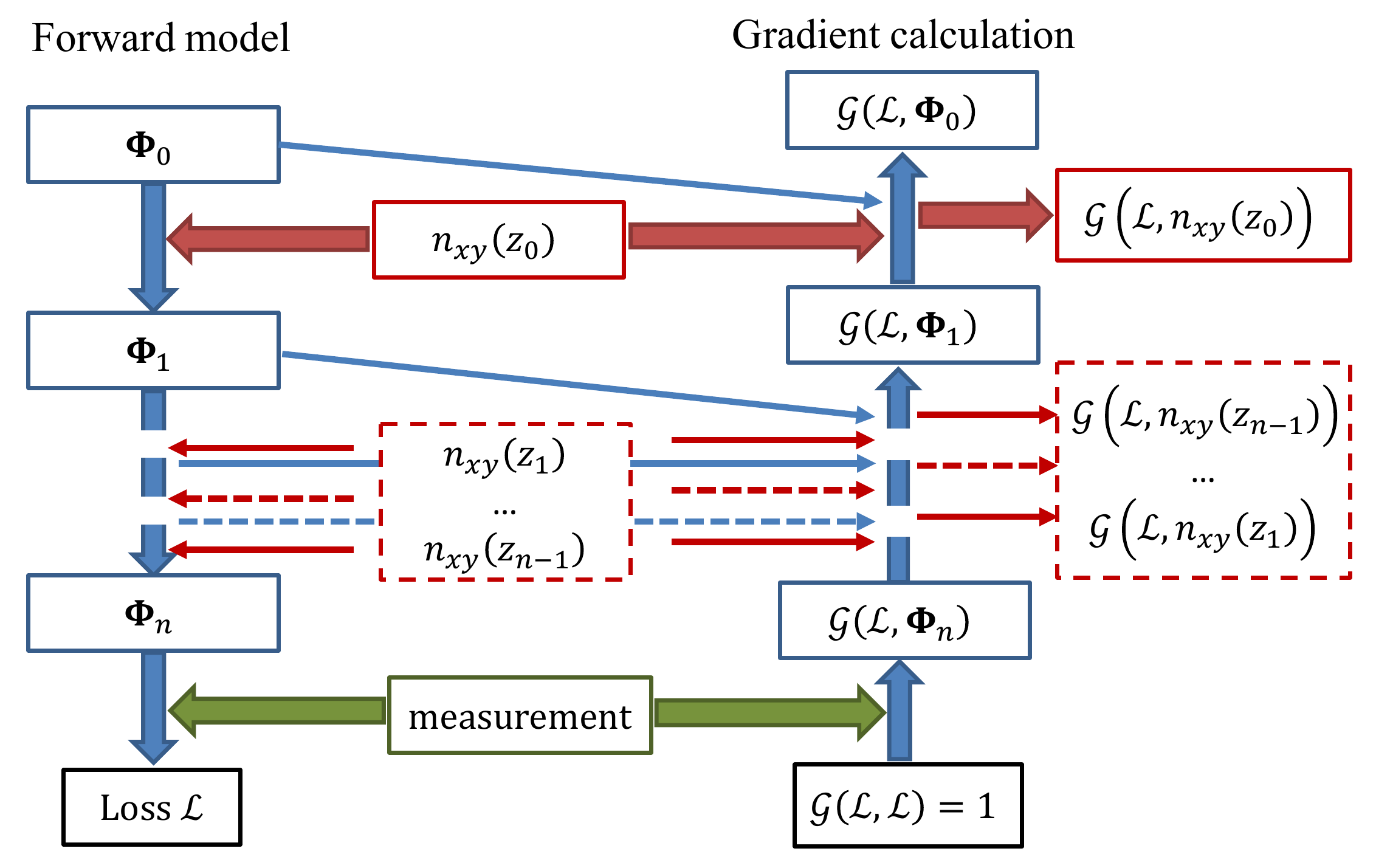}
  \caption{
    Data-flow diagram for forward and gradient calculation.
  }\label{fig2}
\end{figure}

Figure~\ref{fig2} shows the dependency relationship of the data for forward and gradient calculation. In the original automatic differentiation algorithm, all the intermediate fields $\vPhi_{xy}(z_i)$ need to be saved for the gradient calculation during the forward calculation.
For example, to reconstruct a 3D sample discretized to \num{1024 x 1024 x 1024} voxels, the intermediate fields takes \SI{32}{\gibi\byte} memory, which cannot fit into the device memory of most consumer GPUs.
It is possible to use the host memory or the hard disk to store these arrays and transfer the data back and forth between the device memory and the host memory, but this will significantly slow down the reconstruction speed.
Here, we developed a new automatic differentiation framework to significantly reduce the memory consumption of the intermediate field at a slight cost of computing speed.

The SSNP forward model is basically a chain of operators.
If we store the intermediate field at a certain point of the chain, we can resume the algorithm from this point and compute the following slices.
Thus, we can keep only a small portion of the intermediate fields at a few key points, and recompute the others when needed.
A highly efficient strategy is to choose the key points with arithmetic sequence gaps.
For a sample discretized to $n$ slices axially, the number of intermediate fields saved in memory can decrease from $n$ to $m=\left\lceil\frac{\sqrt{1+8n}-1}{2}\right\rceil$, with less than $2\times$ forward computation time.
For the same 3D sample discretized to \num{1024 x 1024 x 1024} voxels, the intermediate field will only take \SI{1.41}{\gibi\byte} memory.
The operator $\lceil\cdot\rceil$ denotes rounding up to an integer.

The details of this strategy are as follows.
When computing the forward model, we select the first field $\vPhi_{xy}(z_1)$ as a key point and drop the following $m-1$ fields.
Then, we save the $(m+1)$th field $\vPhi_{xy}(z_{m+1})$ and drop the following $m-2$ fields.
We do this iteratively until reaching the final slice $\vPhi_{xy}(z_n)$.
During the reconstruction, if one intermediate field does not exist in memory, we resume the forward model from the nearest key point before this slice, recalculate the fields between them, and store all these fields in memory.

It can be proved that the number of fields stored in memory at the same time will never exceed $m$.
We only need to consider the case when $\frac{\sqrt{1+8n}-1}{2}$ is an integer.
Otherwise, the problem is only a sub-problem of a larger $n$ with the same $m=\left\lceil\frac{\sqrt{1+8n}-1}{2}\right\rceil$ and can be solved in the same way.
When $\frac{\sqrt{1+8n}-1}{2}$ is an integer, we have $m=\frac{\sqrt{1+8n}-1}{2}$ or equivantly, $n=\frac{m(m+1)}{2}$.
Since there is one key point per $m, m-1, \dots, 1$ slices, keeping $m$ slices in memory is enough for forward computation.
When doing reconstruction, we compute from the last slice and do the ``gradient back-propagation''.
After we get $\Grad(\Loss,n_{xy}(z_n))$, the intermediate field $\vPhi_{xy}(z_n)$ will never be used later and can be put in a memory pool.
When we compute the $(n-1)$th slice $\Grad(\Loss,n_{xy}(z_{n-1}))$, we can recompute $\vPhi_{xy}(z_{n-1})$ from $\vPhi_{xy}(z_{n-2})$ and save the field using the memory we put in the memory poll before.
The gap between the $i$th key point and the $(i+1)$th key point is $m-i$, and the number of free memory of the used slices exists in the memory pool is also $m-i$, which can hold all the intermediate fields.
Finally, we can successfully compute all the $\Grad(\Loss,n_{xy}(z_i))$ keeping at most $m$ intermediate fields in memory.
Since each slice will be recomputed at most 1 time, the total computation time for forward model is less than 2$\times$ of the original algorithm.
In practice, the gradient computation takes similar time to the forward model computation, and the proximal computation of the TV regularizer also takes about $0.5\times$ of the forward model.
The total time consumption will be $\sim1.4\times$ of the original algorithm.

\begin{figure}[htbp]
  \centering
  \includegraphics[width=0.8\textwidth]{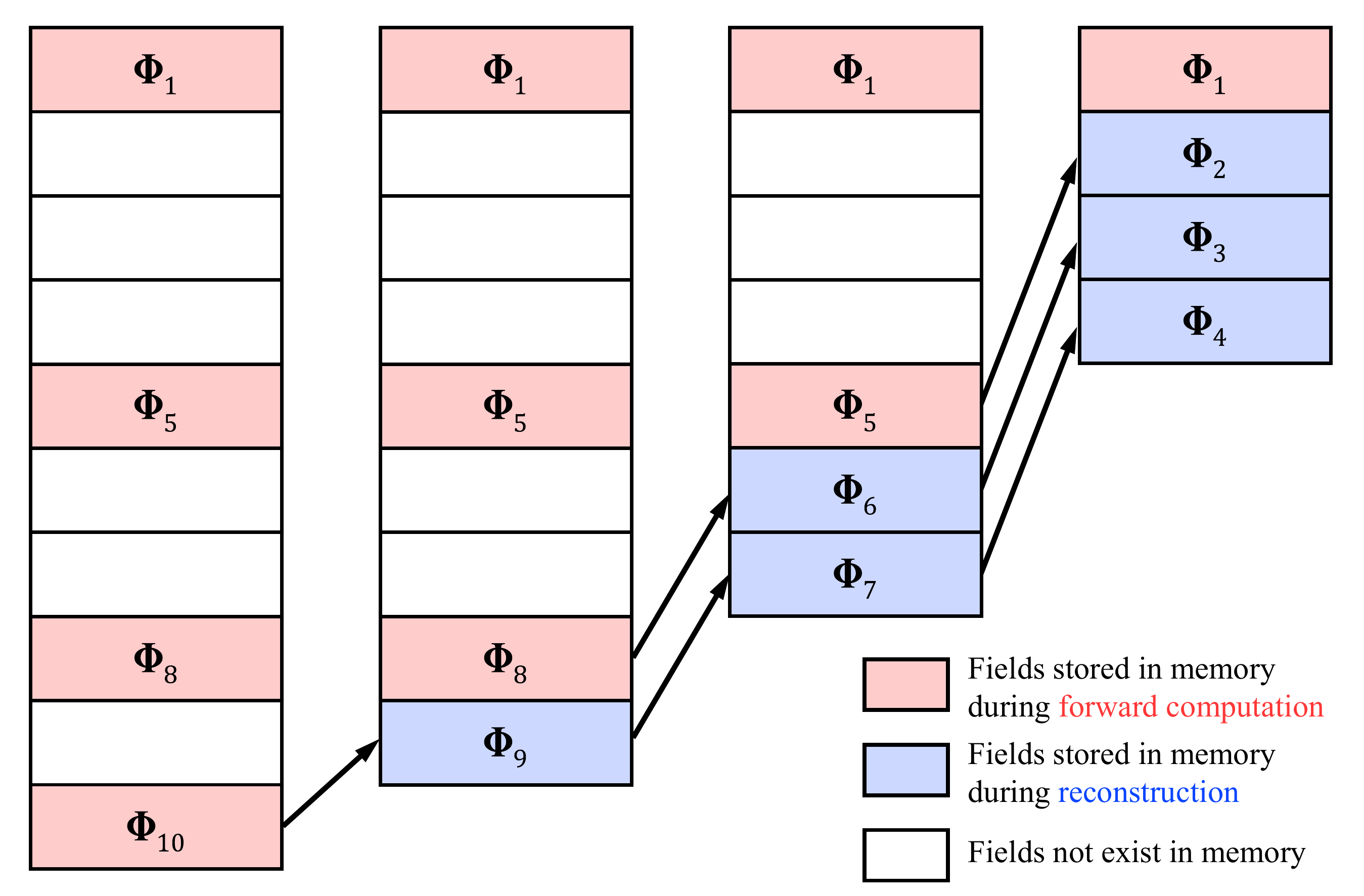}
  \caption{
    Example of our memory management strategy in different reconstruction stage when $n=10$ and $m=\left\lceil\frac{\sqrt{1+8n}-1}{2}\right\rceil=4$.
  }\label{fig2p}
\end{figure}

A summary of this memory-efficient automatic differentiation algorithm is shown in Algorithm~\ref{algo1}, and an  example is shown in Fig.~\ref{fig2p}.

\begin{algorithm}
  \caption{Memory-Efficient Automatic Differentiation}\label{algo1}
  \begin{algorithmic}[1]
    \INPUT{current estimation of the RI distribution $\hat{n}_{xy}$, input field $\vPhi_{xy}(z_0)$ and measurement $I_\mathrm{meas}$}
    \OUTPUT{gradient of RI distribution $\Grad(\Loss, \hat{n}_{xy}$)}
    \State{$m \gets \left\lceil\frac{\sqrt{1+8n}-1}{2}\right\rceil$}
    \State{$r \gets 0$}
    \For{$i \gets 1$ to $n$}
      \State{$\vPhi_{xy}(z_i) \gets \vP\,\vQ(z_i)\,\vPhi_{xy}(z_{i-1})$}
      \State{$r \gets r-1$}
      \If{$r \le 0$}
        \State{store $\vPhi_{xy}(z_i)$ in memory}
        \State{$m \gets m-1$}
        \State{$r \gets m$}
      \EndIf
    \EndFor
    \State{calculate the output field and total intensity on camera plane}
    \State{calculate the gradient $\Grad_n=\Grad(\Loss, \vPhi_{xy}(z_n))$}
    \For{$i \gets n-1$ down to $0$}
      \State{calculate $\Grad'_i$ from $\Grad_{i+1}$}
      \If{$\vPhi_{xy}(z_i)$ is not saved in memory}
        \State{repeat step 4 to calculate $\vPhi_{xy}(z_i)$ from the nearest previous field $\vPhi_{xy}(z_s)$ ($s<i$) and save all the intermediate field in memory}
      \EndIf
      \State{calculate $\Grad(\Loss,\hat{n}_{xy}(z_i))$}
    \EndFor
  \end{algorithmic}
\end{algorithm}

\section{Computational performance evaluation}
The computational implementation plays an important role in this work.
We test our code on 2 different platforms.
Platform 1 is a workstation with one Intel Xeon CPU E5-1630 v4 and one NVIDIA GeForce RTX 2070 GPU, running Windows 10 operating system (version 20H2).
Platform 2 is a shared server node on Boston University Shared Computing Cluster, with 1 core of Inten Xeon CPU E5-2680 v4 and one NVIDIA Tesla P100 GPU (12GB memory, PCIe version), running CentOS Linux operating system (release 7.9.2009).

\begin{table}[htbp]
\centering
\caption{Computation time of each step with GPU and CPU at different platforms (unit: s)}
\begin{tabular}{c|ccccc}
\hline
Processor & Step 1 & Step 2 & Step 3 & Step 4 & Step 5\\
\hline
Platform 1, GPU & 16.4 & 33.7 & 68.6 & 35.1 & 22.9 \\
Platform 2, GPU & 5.5 & 15.3 & 32.0 & 30.6 & 17.3 \\
Platform 1, CPU & 2.2 & 1593.6 & 1708.8 & N.A.$^*$ & 18.1 \\
\hline
\end{tabular}
  \label{tabcomp}

  \begin{flushleft}* We have not implemented TV regularization with FISTA on CPU.\end{flushleft}
\end{table}
\begin{figure}[htbp]
  \centering
  \includegraphics[width=0.8\textwidth]{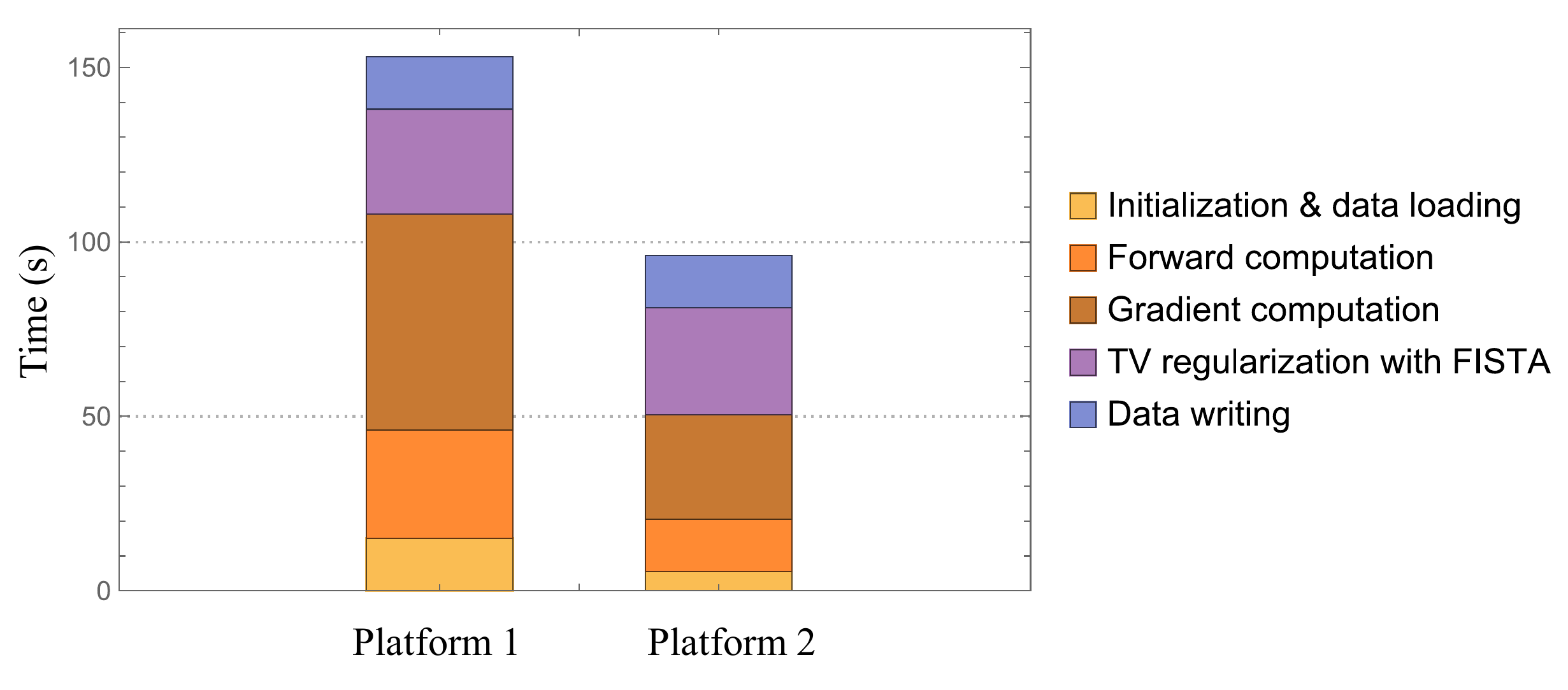}
  \caption{
    Computation time of each step with GPU at different platforms
  }\label{figcomp}
\end{figure}

Table~\ref{tabcomp} and Figure~\ref{figcomp} shows the computational time of all steps of the code.
The details of each step is discussed below:
\begin{enumerate}
  \item Initialization and data loading:
    This step includes loading the library, runtime compilation of the CUDA source code, reading the raw intensity images, and basic preprocessing of the input data.
    Only a small portion of the total computation time should be spend on this part.
    However, due to some unknown problem of PyCUDA library, the CUDA code takes a longer time (\SIrange{15}{30}{\s}) to compile on Windows platform.
  \item Forward computation:
    This step is one core part of the SSNP-IDT model, which computes the intensity estimation and its difference with intensity measurement from the input illumination and the estimated 3D RI distribution.
  \item Gradient computation:
    This step is the other core part of the SSNP-IDT model, which computes the gradient of 3D RI estimation from the loss of previous step.
    The computational complexity of this step is similar to forward computation step when all the intermediate value is saved.
    Here, we use the memory-efficient strategy discussed in previous section to fit this large reconstruction task in GPU memory, which approximately doubles the computation time as a trade-off.
  \item TV regularization with FISTA\@:
    This is an optional step for our iterative SSNP-IDT model when the noise in input data is moderate.
    Although the computational complexity of this step is much smaller than the forward and gradient computation steps, it practically takes similar computation time due to the unoptimized implementation.
  \item Data writing:
    The time cost of this step is mainly decided by the write speed of the hard drive.
    Since we reconstruct a large 3D volume from a limited number of 2D acquisitions, normally the data writing time is much longer than the data loading time.
\end{enumerate}

The performance specification for double precision computation is 203 GFLOPS on NVIDIA GeForce RTX 2070 and 4.7 TFLOPS on NVIDIA Tesla P100.
However, our code cannot fully ultilize the computing power of Tesla P100, which only achieve $\sim 2 \times$ speed-up comparing with RTX 2070.
The speed may be limited by memory bandwidth and the overhead of loading CUDA functions synchronously with Python.

In conclusion, our code can achieve $30 \sim 60 \times$ acceleration ratio on different GPUs comparing to single-thread CPU version.
This enables us to perform SSNP-IDT reconstruction for dynamic samples in a resonable time.
